\documentclass[]{spie}

\pdfoutput=1
\usepackage{amsmath,amsfonts,amssymb}
\usepackage{graphicx}
\usepackage[colorlinks=true, allcolors=blue]{hyperref}
\usepackage{booktabs, multirow}
\usepackage{textcomp}
\usepackage{adjustbox}
\usepackage{import}

\title{Broadband anti-reflective coatings for cosmic microwave background experiments}

\author[a]{A.~Nadolski}
\author[a]{A.~M.~Kofman}
\author[a,b]{J.~D.~Vieira}
\author[c]{P.~A.~R.~Ade}
\author[d,e]{Z.~Ahmed}
\author[f,g]{A.~J.~Anderson}
\author[h]{J.~S.~Avva}
\author[g]{R.~Basu Thakur}
\author[i,g]{A.~N.~Bender}
\author[f,g,j]{B.~A.~Benson}
\author[g,k,l,i,j]{J.~E.~Carlstrom}
\author[i,g]{F.~W.~Carter}
\author[i]{T.~W.~Cecil}
\author[i,g,j]{C.~L.~Chang}
\author[m]{J.~F.~Cliche}
\author[h]{A.~Cukierman}
\author[h]{T.~de~Haan}
\author[n]{J.~Ding}
\author[m,o]{M.~A.~Dobbs}
\author[g,l]{D.~Dutcher}
\author[p]{W.~Everett} 
\author[q]{A.~Foster} 
\author[a]{J.~Fu} 
\author[g,r]{J.~Gallicchio} 
\author[m]{A.~Gilbert} 
\author[h]{J.~C.~Groh} 
\author[h]{S.~T.~Guns} 
\author[a]{R.~Guyser} 
\author[p,s]{N.~W.~Halverson} 
\author[a,i]{A.~H.~Harke-Hosemann} 
\author[h]{N.~L.~Harrington} 
\author[g]{J.~W.~Henning} 
\author[h]{W.~L.~Holzapfel} 
\author[h]{N.~Huang} 
\author[d,t,e]{K.~D.~Irwin} 
\author[h]{O.~B.~Jeong} 
\author[f]{M.~Jonas} 
\author[u]{A.~Jones} 
\author[n]{T.~S.~Khaire} 
\author[q]{M.~Korman} 
\author[f]{D.~L.~Kubik} 
\author[i]{S.~Kuhlmann} 
\author[d,t,e]{C.-L.~Kuo} 
\author[h,v]{A.~T.~Lee} 
\author[g]{A.~E.~Lowitz} 
\author[g,k,l,j]{S.~S.~Meyer} 
\author[u]{D.~Michalik} 
\author[m]{J.~Montgomery} 
\author[w]{T.~Natoli} 
\author[f]{H.~Nguyen} 
\author[m]{G.~I.~Noble} 
\author[n]{V.~Novosad} 
\author[g]{S.~Padin} 
\author[g,l]{Z.~Pan} 
\author[n]{J.~Pearson} 
\author[n]{C.~M.~Posada} 
\author[g,l]{W.~Quan} 
\author[f,g]{A.~Rahlin} 
\author[q]{J.~E.~Ruhl} 
\author[p]{J.T.~Sayre} 
\author[g,j]{E.~Shirokoff} 
\author[x]{G.~Smecher} 
\author[g,l]{J.~A.~Sobrin} 
\author[y]{A.~A.~Stark} 
\author[d,t]{K.~T.~Story} 
\author[v]{A.~Suzuki}
\author[d,t,e]{K.~L.~Thompson}
\author[c]{C.~Tucker}
\author[w,z]{K.~Vanderlinde}
\author[i]{G.~Wang}
\author[aa,h]{N.~Whitehorn}
\author[i]{V.~Yefremenko}
\author[d,t,e]{K.~W.~Yoon}
\author[z]{M.~R.~Young}

\affil[a]{Department of Astronomy, University of Illinois at Urbana-Champaign, 1002 West Green Street, Urbana, IL, USA 61801}
\affil[b]{Department of Physics, University of Illinois Urbana-Champaign, 1110 West Green Street, Urbana, IL, USA 61801}
\affil[c]{School of Physics and Astronomy, Cardiff University, Cardiff CF24 3YB, United Kingdom}
\affil[d]{Kavli Institute for Particle Astrophysics and Cosmology, Stanford University, 452 Lomita Mall, Stanford, CA, USA 94305}
\affil[e]{SLAC National Accelerator Laboratory, 2575 Sand Hill Road, Menlo Park, CA, USA 94025}
\affil[f]{Fermi National Accelerator Laboratory, MS209, P.O. Box 500, Batavia, IL, USA 60510}
\affil[g]{Kavli Institute for Cosmological Physics, University of Chicago, 5640 South Ellis Avenue, Chicago, IL, USA 60637}
\affil[h]{Department of Physics, University of California, Berkeley, CA, USA 94720}
\affil[i]{High-Energy Physics Division, Argonne National Laboratory, 9700 South Cass Avenue., Argonne, IL, USA 60439}
\affil[j]{Department of Astronomy and Astrophysics, University of Chicago, 5640 South Ellis Avenue, Chicago, IL, USA 60637}
\affil[k]{Enrico Fermi Institute, University of Chicago, 5640 South Ellis Avenue, Chicago, IL, USA 60637}
\affil[l]{Department of Physics, University of Chicago, 5640 South Ellis Avenue, Chicago, IL, USA 60637}
\affil[m]{Department of Physics and McGill Space Institute, McGill University, 3600 Rue University, Montreal, Quebec H3A 2T8, Canada}
\affil[n]{Materials Sciences Division, Argonne National Laboratory, 9700 South Cass Avenue, Argonne, IL, USA 60439}
\affil[o]{Canadian Institute for Advanced Research, CIFAR Program in Cosmology and Gravity, Toronto, ON, M5G 1Z8, Canada}
\affil[p]{CASA, Department of Astrophysical and Planetary Sciences, University of Colorado, Boulder, CO, USA 80309}
\affil[q]{Department of Physics, Center for Education and Research in Cosmology and Astrophysics, Case Western Reserve University, Cleveland, OH, USA 44106}
\affil[r]{Harvey Mudd College, 301 Platt Boulevard., Claremont, CA, USA 91711}
\affil[s]{Department of Physics, University of Colorado, Boulder, CO, USA 80309}
\affil[t]{Department of Physics, Stanford University, 382 Via Pueblo Mall, Stanford, CA, USA 94305}
\affil[u]{University of Chicago, 5640 South Ellis Avenue, Chicago, IL, USA 60637}
\affil[v]{Physics Division, Lawrence Berkeley National Laboratory, Berkeley, CA, USA 94720}
\affil[w]{Dunlap Institute for Astronomy \& Astrophysics, University of Toronto, 50 St. George Street, Toronto, ON, M5S 3H4, Canada}
\affil[x]{Three-Speed Logic, Inc., Vancouver, B.C., V6A 2J8, Canada}
\affil[y]{Harvard-Smithsonian Center for Astrophysics, 60 Garden Street, Cambridge, MA, USA 02138}
\affil[z]{Department of Astronomy \& Astrophysics, University of Toronto, 50 St. George Street, Toronto, ON, M5S 3H4, Canada}
\affil[aa]{Department of Physics and Astronomy, University of California, Los Angeles, CA, USA 90095}

\authorinfo{Further author information: (Send correspondence to Andrew Nadolski)\\
E-mail: nadolsk1@illinois.edu}

\begin{document} 
\maketitle

\begin{abstract}
The desire for higher sensitivity has driven ground-based cosmic microwave background (CMB) experiments to employ ever larger focal planes, which in turn require larger reimaging optics.
Practical limits to the maximum size of these optics motivates the development of quasi-optically-coupled (lenslet-coupled), multi-chroic detectors.
These detectors can be sensitive across a broader bandwidth compared to waveguide-coupled detectors.
However, the increase in bandwidth comes at a cost: the lenses (up to $\sim$700 mm diameter) and lenslets ($\sim$5 mm diameter, hemispherical lenses on the focal plane) used in these systems are made from high-refractive-index materials (such as silicon or amorphous aluminum oxide) that reflect nearly a third of the incident radiation.
In order to maximize the faint CMB signal that reaches the detectors, the lenses and lenslets must be coated with an anti-reflective (AR) material.
The AR coating must maximize radiation transmission in scientifically interesting bands and be cryogenically stable.
Such a coating was developed for the third generation camera, SPT-3G, of the South Pole Telescope (SPT) experiment, but the materials and techniques used in the development are general to AR coatings for mm-wave optics.
The three-layer polytetrafluoroethylene-based AR coating is broadband, inexpensive, and can be manufactured with simple tools.
The coating is field tested; AR coated focal plane elements were deployed in the 2016-2017 austral summer and AR coated reimaging optics were deployed in 2017-2018.
\end{abstract}

\keywords{Anti-reflective coating, broadband, millimeter-wave}

\section{INTRODUCTION}
\label{sec:intro}

The cosmic microwave background (CMB) encodes a wealth of information about our universe: the temperature anisotropy tells us about primordial density fluctuations \cite{barrow1991}; measurements of the temperature power spectrum indicate the curvature and composition of the early universe \cite{bernardis2000, planck2016_part13}; the polarization anisotropy gives us information about the energy scale of inflation \cite{hu2003}.
Detailed measurements of CMB properties serve to further our understanding of fundamental questions in cosmology and astrophysics.
Demand for higher sensitivity has pushed the CMB community to design experiments with large optical elements and large focal planes.
One such experiment is the South Pole Telescope's (SPT) third generation camera, called SPT-3G. 

SPT-3G, deployed in the austral summer of 2016-2017, is designed to achieve a roughly order-of-magnitude increase in mapping speed over its predecessor thanks to its wide-field optics that illuminate the entire focal plane and its broadband pixels \cite{benson2014}, which are a departure from optical chain and detector coupling designs of previous CMB experiments \cite{keating2003, chang2009, essinger2010, niemack2010, austermann2012, tran2009, hanany2013}.
As current CMB detector technology is sensitvity-limited by photon shot noise, one way to boost an experiment's overall sensitivity is to increase the size of the focal plane (i.e., the total number of detectors).
There is, however, a practical upper limit to the area of a focal plane: for the materials and processes currently used in the CMB field, it becomes difficult (and expensive) to manufacture high-refractive-index lenses as they approach $\sim$1 m diameter.
Rather than increasing the physical area of the focal plane, another way to increase overall sensitivity is by increasing the bandwidth of the focal plane.
Multi-chroic detectors (i.e., detectors sensitive to more than one bandpass, or color) include multiple colors within the same pixel footprint, which allows for efficient use of the limited focal plane area by increasing the total bandwidth of each pixel.

SPT-3G employs high-refractive-index reimaging lenses and microfabricated arrays of polarization-sensitive, multi-chroic pixels on the focal plane.
Each pixel consists of a broadband, planar, log-periodic sinuous antenna, sensitive to two linear polarizations of light.
Each polarization element is coupled to transition edge sensors (TES) tuned to 95, 150, and 220 GHz with $\sim$30\% bandwidth in each spectral band \cite{suzuki2018, posada2015}.
In total, there are six bolometers per pixel.
The multi-chroic nature of each individual pixel translates to greater total bandwidth on the focal plane than achieved in previous CMB experiments.
Additionally, SPT-3G's use of high-refractive-index, wide-field optics allows for a large focal plane area, which in turn allows for a large absolute number of detectors.

\begin{figure}[b]
\begin{center}
\begin{tabular}{c}
\includegraphics[height=5.5cm]{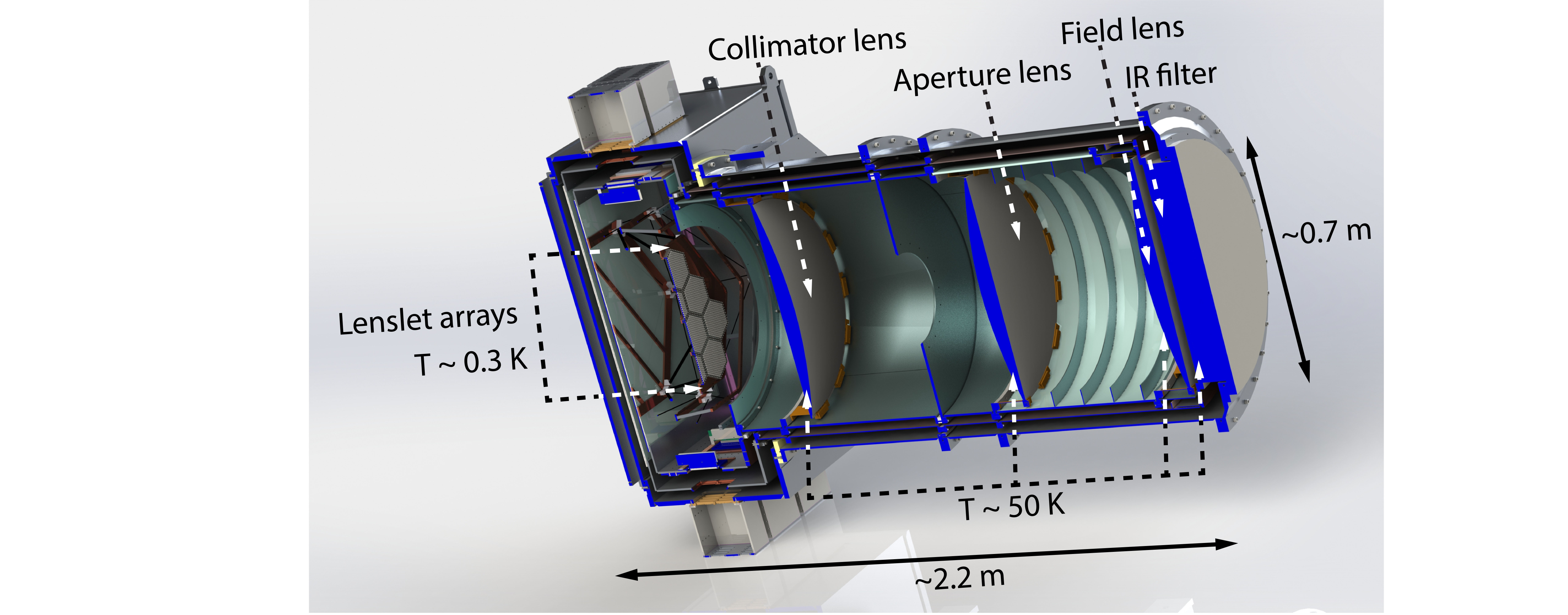}
\end{tabular}
\end{center}
\caption[example]
{ \label{fig:cryostat}
A section-view of the SPT-3G cryostat. The cylindrical portion houses the large-format optics (infrared filter; field, aperture, and collimator lenses; T $\sim$ 50 K), and the box houses the focal plane (lenslet arrays; T $\sim$ 0.3 K).
}
\end{figure}

In the case of SPT-3G, the wide-field optics are $\sim$720 mm in diameter and made from amorphous aluminum oxide (Al$_{2}$O$_{3}$; henceforth, alumina).
There are three alumina lenses and one alumina infrared filter in the SPT-3G optical chain; the final focusing elements (arrays of 5 mm diameter hemispherical lenslets) are also made from alumina (Figure \ref{fig:cryostat}).
Due to the high refractive index of alumina ($n\approx3$), nearly 30\% of incident radiation is reflected at each alumina surface.
Since CMB detectors are sensitivity-limited by photon shot noise, the reflection and scattering of CMB photons away from detectors has a direct impact on the overall sensitivity of the experiment.
Furthermore, increased reflection leads to increased thermal loading throughout the interior of the experiment, which can in turn lead to challenges in managing the cryogenic environment.
Consequently, minimizing loss due to reflections is a high priority.

Single-layer and two-layer anti-reflective (AR) coatings that provide good single- and dual-band coverage have been developed and demonstrated \cite{quealy2012, hargrave2010, rosen2013, pb2_instrument2016, wheeler2014}.
In this work we present a three-layer polytetrafluoroethylene-based (PTFE) millimeter wave AR coating, developed for the SPT-3G camera and operational in the field since 2016.
In Sec. \ref{sec:design} we address the design requirements for the SPT-3G AR coating;
in Sec. \ref{sec:fabrication} we describe how the coating is fabricated.
We present measurements of the coating's performance in Sec. \ref{sec:characterization}, discuss our results in Sec. \ref{sec:discussion}, and draw conclusions in Sec. \ref{sec:conclusion}.

\section{AR COATING DESIGN}
\label{sec:design}

The initial development and materials selection for the SPT-3G AR coating were made in the context of the focal plane's lenslet arrays (271-element arrays composed of 5 mm diameter, hemispherical alumina lenses), the final element in the optical chain.
After the initial deployment of SPT-3G, we adapted the AR coating materials and the techniques we had developed to the large-format lenses for the second season of operation.
When designing the SPT-3G AR coating, we required that the end product must: 1) provide AR coverage from $\sim$80 GHz to $\sim$260 GHz; 2) conform to high-curvature surfaces; 3) be able to be fabricated en masse, rather than for each optical element individually.

A perfect single frequency coating AR coating is one quarter-wavelength thick at the target frequency, and has a refractive index equal to the geometric mean of the incident and terminal media refractive indices: $n_{AR}=\sqrt{n_{i}n_{f}}$.
This type of coating is the thinnest possible configuration for a given frequency and AR coating material.
Minimizing the thickness of an AR coating is important because physical materials are lossy.
The refractive index of physical media are generally complex-valued, with the non-zero imaginary part describing electromagnetic attenuation within those media -- and dielectric loss is proportional to material thickness (and frequency).
Conversely, a (nearly) perfect broadband coating is one that approximates a Klopfenstein taper \cite{klopfenstein1956}, a well-known microwave transmission line impedance transformer.
The analogous AR coating is one that gradually and continuously transitions from the refractive index of the incident medium to the refractive index of the terminal medium.
In the case of SPT-3G, such a coating would vary smoothly and continuously from $n=1$, the refractive index of vacuum, to $n\approx3$, the refractive index of alumina (which is similar to that of silicon).
In practice, a Klopfenstein-approximate coating would not only be difficult to manufacture, but it would also be thick.
A thick coating, while perhaps providing better AR properties, will cause more attenuation than a thinner coating.

An intermediate configuration composed of multiple discrete layers, each with a distinct refractive index, is a compromise between narrow AR coverage with low loss and broad AR coverage with high loss.
Commercial thermoplastic sheets are well-suited to this intermediate configuration.
The electrical properties and physical dimensions of thermoplastics used in microwave engineering applications are strictly controlled, making it easier to create a consistent, layered coating.
In addition, thermoplastics have previously been used as single-layer coatings for high-curvature lenslets \cite{quealy2012}.

Previous CMB AR coating development \cite{suzuki2013} determined that our ideal AR coating should have three layers, and that each discrete layer should have refractive index 1.41, 2.0, and 2.65 and thickness 331, 234, and 171 $\mu$m, respectively.
These refractive indices and thicknesses comprise an AR coating optimized for transmission at 160 GHz that has a broad enough response to cover our low (95 GHz) and high (220 GHz) bands.
However, commercially available thermosplastics with arbitrary refractive index and thickness are not readily available.
Therefore, we chose materials with characteristics as close to the ideal values as possible.

We found that commercially available PTFE-based products most closely satisfied our refractive index and curvature requirements.
In its raw state, PTFE is a powder which can be cast and sintered, or extruded to produce a wide range of shapes.
Because of its favorable properties (e.g., low dielectric loss and coefficient of friction, hydrophobicity, and thermal stability) PTFE has seen widespread use in communications and microwave engineering applications.
The electrical properties of raw PTFE can be modified by changes to the powder mixture before sintering (e.g., additives) and other processes.
Each layer comprising the SPT-3G AR coating has undergone some form of additional processing.

For the low-index layer we chose porous PTFE (Porex for lenslet arrays and Zitex for large-format optics), while for the middle- and high-index layers we chose filled PTFE (Rogers Corporation RO3035 bondply and Rogers Corporation RO3006 bondply, respectively).
Porous PTFE has a lower refractive index than pure, non-porous PTFE, whereas filled PTFE (which often includes additives such as high-index ceramics or glass) have a higher refractive index than pure PTFE.
Table \ref{table:materials} contains details of the properties of each AR layer.
The materials we chose are available as 18 x 24 x 0.005 inch sheets (RO3035 and RO3006), 13 x 0.015 inch rolls of arbitrary length (Porex), and 10 x 0.015 inch rolls of arbitrary length (Zitex).
Zitex, which has similar properties to Porex, was chosen for the large-format optics because it is less expensive and has a shorter manufacturing lead time.

Using the transfer matrix method \cite{hou1974}, we simulated transmission for a range of layer configurations and found that one stock-thickness layer of each material was an acceptable compromise between coating thickness and overall transmission.
The thickness constraint of the coating was driven mainly by the pitch of the detector pixels: 6.789 mm.
A coating made too thick would cause adjacent pixels to physically interfere with each other.
Due to their different geometries, the lenslet and large-format AR coating fabrication processes vary in how the primary materials are made to adhere to each other and to the optical element.
We describe the two fabrication processes below.

\begin{table}[tb]
\caption{Select properties of AR coating, lens, and lenslet materials.
With the exception of Zitex, the values given are those given as product specifications by the manufacturers\cite{porex2015, benford2003, rogers2015, coorstek2016}.
Column four indicates the frequency at which the loss tangent and refractive index were measured.
Columns five and six give the loss tangent and refractive index, respectively.
Column seven denotes the type of AR coating in which the material was used, and column eight gives the bibliography entry for convenience.}
\label{table:materials}
\begin{center}
\begin{tabular}{lllccrlr}
\hline
\multirow{1}{*}{Common Name} & \multirow{1}{*}{Trade Name} & \multirow{1}{*}{Manufacturer} &\multirow{1}{*}{$\nu_{test}$ [GHz]} & \multirow{1}{*}{$\tan\delta [\times 10^{-3}$]} & \multirow{1}{*}{$n$} & \multirow{1}{*}{Coating type}  & \multirow{1}{*}{Refs.}\\
\hline
Porous PTFE  &  PM-23J     &  Porex         &  10   &  0.6        &  1.32     &  Lenslet       &  \citenum{porex2015}     \\
Porous PTFE  &  G-115      &  Zitex         &  400  &  $\sim$0.6  &  $>$1.20  &  Large-format  &  \citenum{benford2003}   \\
Filled PTFE  &  RO3035     &  Rogers Corp.  &  10   &  1.5        &  1.87     &  Both          &  \citenum{rogers2015}    \\
Filled PTFE  &  RO3006     &  Rogers Corp.  &  10   &  2.0        &  2.56     &  Both          &  \citenum{rogers2015}    \\
Alumina      &  AD-995-I   &  CoorsTek      &  5    &  0.1        &  3.13     &  Lenslet       &  \citenum{coorstek2016}  \\
Alumina      &  AD-995-LT  &  CoorsTek      &  5    &  0.06       &  3.13     &  Large-format  &  \citenum{coorstek2016}  \\
\hline
\end{tabular}
\end{center}
\end{table}

\section{FABRICATION}
\label{sec:fabrication}
The SPT-3G focal plane is composed of ten modules.
Each module has 271 pixels, and each pixel is quasi-optically coupled by a hemispherical alumina lenslet that is 5 mm in diameter.
Individually AR coating and assembling lenslets was not feasible owing to the large number of lenslets per module and their dense packing (the lenslets are centered on a 6.789 mm pitch; see Fig. \ref{fig:lenslet_array}).
Consequently, we developed a high-throughput coating process that enabled us to create a monolithic three-layer coating (i.e., without any bonding agent between layers) and, as a separate step, apply it to the entire lenslet array at one time.
Table \ref{table:lenslet_timeline} describes the approximate time required per step for lenslet array fabrication.
The large-format optics consist of three reimaging lenses and a flat infrared filter.
The outer dimesion of these parts is $\sim$720 mm.
The process we used for AR coating the large-format optics is essentially an industrial vacuum bagging process.
Table \ref{table:lens_timeline} describes the approximate time required per step for coating large-format optics.
Below, we give a brief overview of each process in turn.

\subsection{Lenslet Coating}
\label{subsec:lenslet_coating}
\begin{figure}[b]
\begin{center}
\begin{tabular}{cc}
\includegraphics[height=5.5cm]{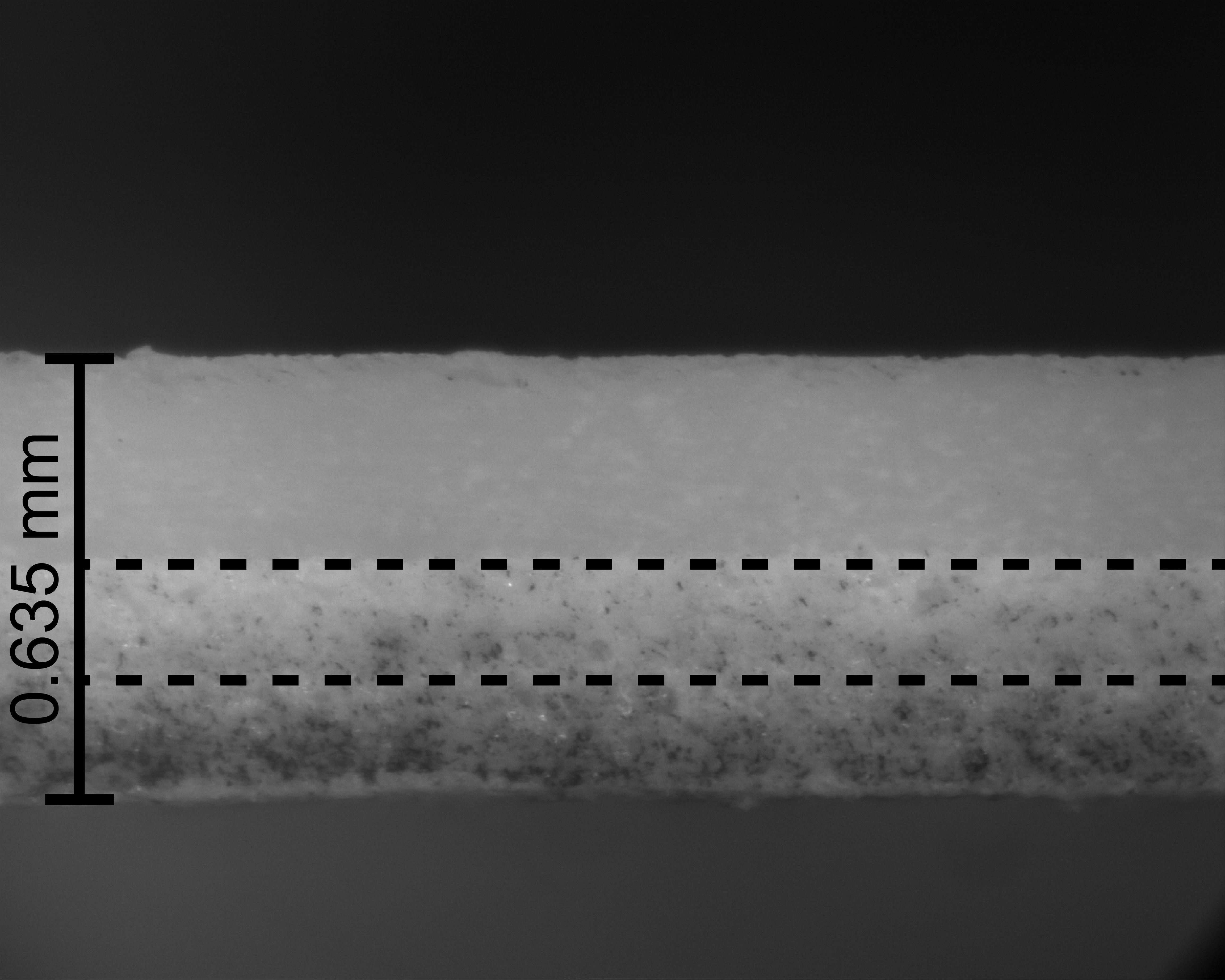}
\includegraphics[height=5.5cm]{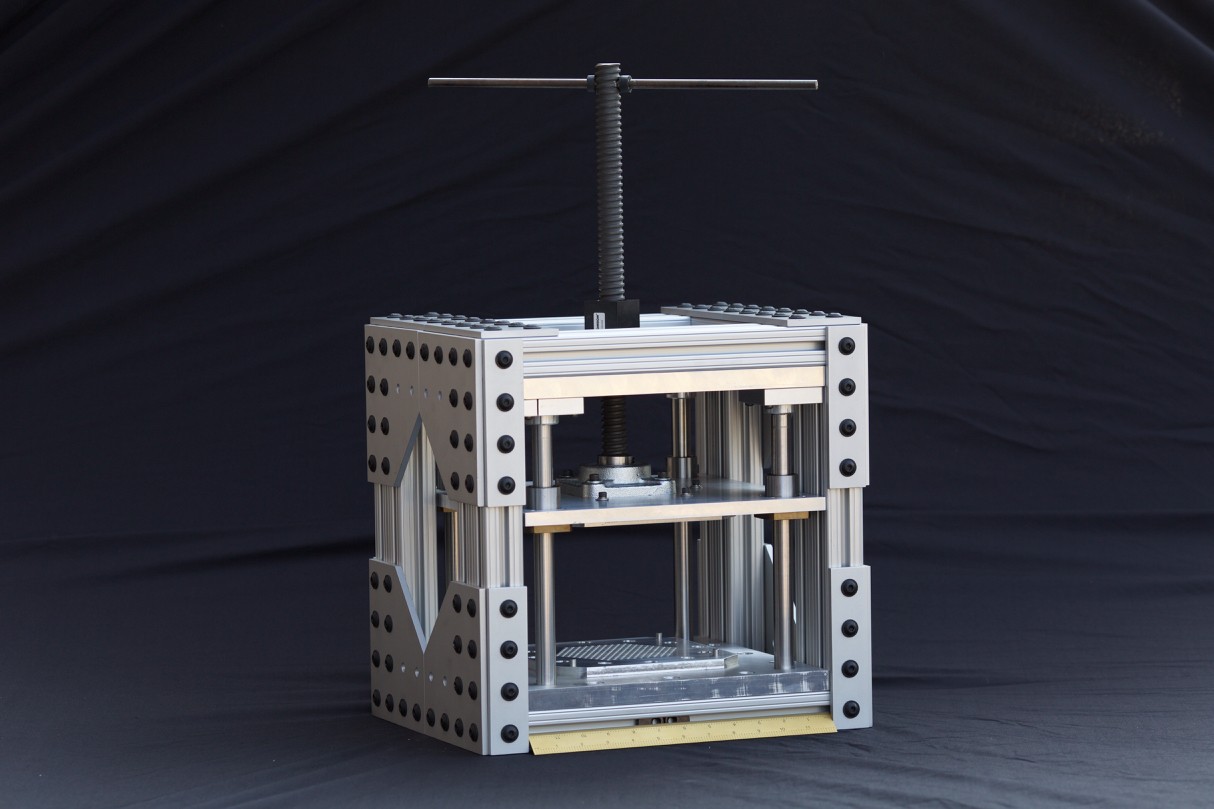}
\end{tabular}
\end{center}
\caption[example]
{ \label{fig:cross_section}
\emph{Left:} A cross-section-view photograph of a laminated, monolithic, PTFE-based, three-layer AR coating as seen in the infrared (scale is approximate). The distinct layers are visible: Porex (top, lowest index), RO3035 (middle, middle index), RO3006 (bottom, highest index). The total thickness of the coating is $\sim$0.635 mm.
\emph{Right:} The screw-driven die press used to form the lenslet AR coating. The gold-colored ruler at the base of the press is for scale.}
\end{figure}

Fabrication of the lenslet AR coating begins with laminating the three different layers together to make a monolithic sheet, roughly 150 mm x 150 mm.
We stack the three separate, PTFE-based sheets in a mold, apply mechanical pressure to that mold, and then heat the entire assembly.
As PTFE approaches its glass transition temperature ($\sim$390\textdegree C) it enters a gel-like state that resists deformation.
Under heat and pressure, the PTFE molecules coalesce and diffuse across the boundaries between the individual sheets.
As the mold and coating is slowly cooled, the coating solidifies into a single sheet.
Figure \ref{fig:cross_section} (left panel) depicts the result of the lamination process.

After the coating is laminated it is shaped to conform to the lenslet array.
To shape the coating, we use a screw-driven die press and system of molds (Fig. \ref{fig:cross_section} right panel).
The laminated sheet is aligned in the mold and compressed until it reaches the appropriate curvature; the result is a sheet that mates directly to the lenslet array.
Following the molding process, the coating is secured to the lenslet array by a thin layer of Stycast 1266 applied to each lenslet; we chose Stycast 1266 for its low viscosity.
The coated lenslet array is then returned to the screw-driven die press and allowed to cure under pressure for at least 12 hours.

The final step in the lenslet AR coating process is laser dicing the coating.
Laser dicing is necessary to relieve thermal stress that would be caused by differential thermal contraction between the PTFE and silicon lenslet wafer.
Without stress relief, a coated lenslet array cooled to $\sim$100K will shatter.
We use a 30 W CO$_{2}$ laser\footnote{Epilogue Zing 16; \url{https://www.epiloglaser.com/laser-machines/zing-laser-series.htm}} to cut a hexagon around each lenslet, which decouples each lenslet from its neighbors and minimizes the total contractile area of any one element (Fig. \ref{fig:lenslet_array}).

\begin{table}[tb]
\caption{An approximate timeline for AR coating a single lenslet array. The ``Step time'' column is divided into ``Hands on'' and ``Process''. Hands on time is that spent physically working on a part, while process time is the time required for a process to complete (e.g., oven cycle or epoxy curing). Note that steps 1 and 2 can be executed in parallel.}
\label{table:lenslet_timeline}
\begin{center}
\begin{tabular}{clrrrr}
\hline
\multirow{2}{*}{Step number} & \multirow{2}{*}{Description} & \multicolumn{2}{c}{Step time [h]} & \multicolumn{2}{c}{Cumulative time [h]} \\
\cline{3-4} \cline{5-6}
& & Hands on & Process & Hands on & Total \\
\hline
1  &  Coating lamination   &  1    &  12   &  1    &  13    \\
2  &  Array preparation    &  1.5  &  12   &  2.5  &  26.5  \\
3  &  Coating application  &  1.5  &  12   &  4    &  40    \\
4  &  Coating dicing       &  1    &  $-$  &  5    &  41    \\
\hline
\end{tabular}
\end{center}
\end{table}

\subsection{Large Format Optics Coating}
Unlike the lenslet coating, where the PTFE layers are bonded directly together, the large optics coating employs a low-density polyethylene (LDPE) bonding layer between AR coating layers.
The intermediate LDPE bonding layers are used in the large-format coatings because it is difficult to make a large enough lamination mold to create monolithic sheets.
LDPE begins to flow at $\sim$110\textdegree C, below the glass transition temperature of PTFE ($\sim$390\textdegree C).
Therefore, the primary AR coating materials remain stable while the LDPE melts and adheres to them (in the case of the innermost layer, the LDPE adheres to the alumina as well).

\begin{figure}[b]
\begin{center}
\begin{tabular}{c}
\includegraphics[height=3cm]{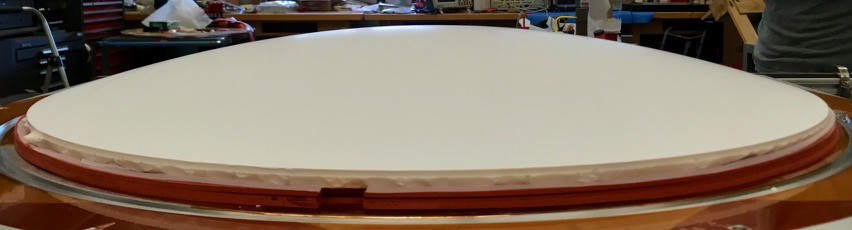}
\end{tabular}
\end{center}
\caption[lens]
{ \label{fig:large_lens}
A photograph of the SPT-3G collimator lens after having been AR coated. The outer diameter of the coated surface is $\sim$700 mm. The outer lip of this lens (a non-optical, though mechanically important part) was damaged during the AR coating process, but we adjusted the vacuum bagging procedure and no additional lenses were damaged.}
\end{figure}

A second reason for using the intermediate LDPE layers is that it becomes difficult to ensure the coating lays flat on a curved surface as the thickness of the coating increases.
We avoid the thickness problem by applying only one AR layer at a time.
This process means that the lens is baked a total of four times: once for each layer and one final baking to ensure that seams in the coating are properly adhered.
Figure \ref{fig:large_lens} shows an AR coated SPT-3G lens, the result of our first end-to-end attempt at the lens AR coating process.

Seams in the AR coating are an inevitable consequence of the form factor of the coating materials; namely, the coating materials are not large enough to cover the large optics as a single sheet.
Therefore, each layer comprises multiple sheets, and the edges of those sheets overlap during baking.
However, the overlapping edges increase the layer thickness at the seams, which is undesirable.
To avoid that extra thickness, we add a temporary strip of pure PTFE between the LDPE layer and the position of the seam, which keeps the seam material from adhering to the LDPE.
The pure PTFE does not adhere well to the LDPE during baking and does not adhere to the AR layer at all.
The temporary strip is removed after the lens is taken out of the oven.
We then remove the excess overlapping material.
We do this by using a razor to make a depth-controlled cut down the middle of the long axis of the seam; we control the depth to avoid cutting any material below the seam.
We then remove the free-floating AR material, leaving the rest of the coating surface flat.
This process is repeated for each layer of the AR coating.
Table \ref{table:lens_timeline} gives a broad overview of the large optics AR coating process.

\begin{table}[t]
\caption{An approximate timeline for AR coating a single side of a large-format optical element. The ``Step time'' column is divided into ``Hands on'' and ``Process''. Hands on time is that spent physically working on a part, while process time is the time required for a process to complete (e.g., oven cycle).}
\label{table:lens_timeline}
\begin{center}
\begin{tabular}{clrrrr}
\hline
\multirow{2}{*}{Step number} & \multirow{2}{*}{Description} & \multicolumn{2}{c}{Step time [h]} & \multicolumn{2}{c}{Cumulative time [h]} \\
\cline{3-4} \cline{5-6}
& & Hands On & Process & Hands On & Total \\
\hline
1  &  Material preparation      &  4    &  $-$ &  4     &  4     \\
2  &  Inner layer application   &  1.5  &  10  &  5.5   &  15.5  \\
3  &  Middle layer application  &  3    &  10  &  8.5   &  28.5  \\
4  &  Outer layer application   &  3    &  10  &  11.5  &  41.5  \\
5  &  Final bake                &  1    &  10  &  12.5  &  52.5  \\
\hline
\end{tabular}
\end{center}
\end{table}

\section{CHARACTERIZATION}
\label{sec:characterization}
The lenslet AR coating was measured by means of a Micheleson interferometer Fourier Transform Spectrometer (FTS) at University of California Berkeley.
The FTS consists of a modulated blackbody radiation source, collimating optics, and a broadband detector \cite{suzuki2013}.
Generally, in FTS measurements the response of the system is measured with a clear detector aperture, then the sample is placed in the aperture and its response is measured.

\begin{figure}[b]
\begin{center}
\begin{tabular}{cc}
\includegraphics[height=5.5cm]{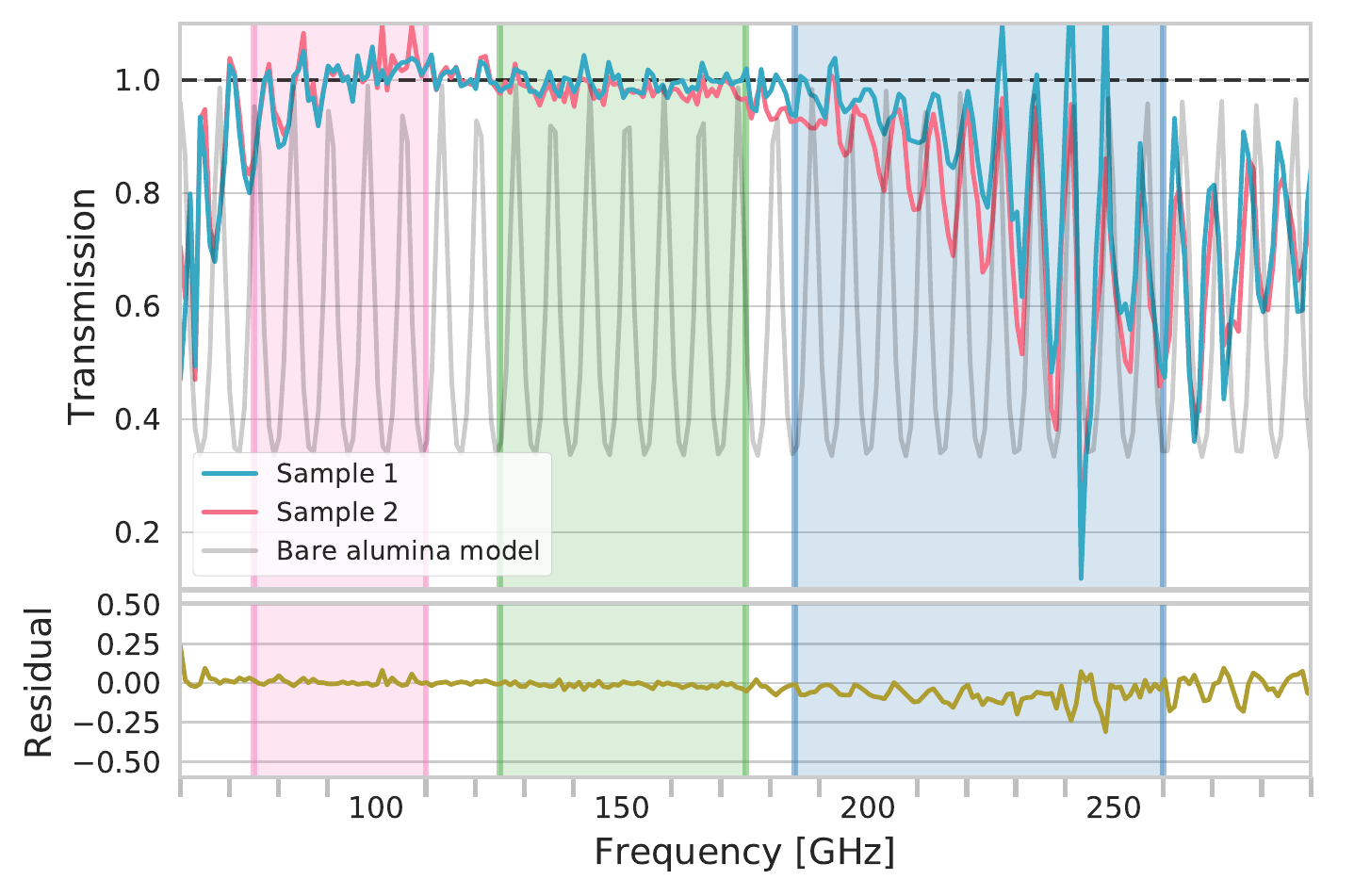}
\includegraphics[height=5.5cm]{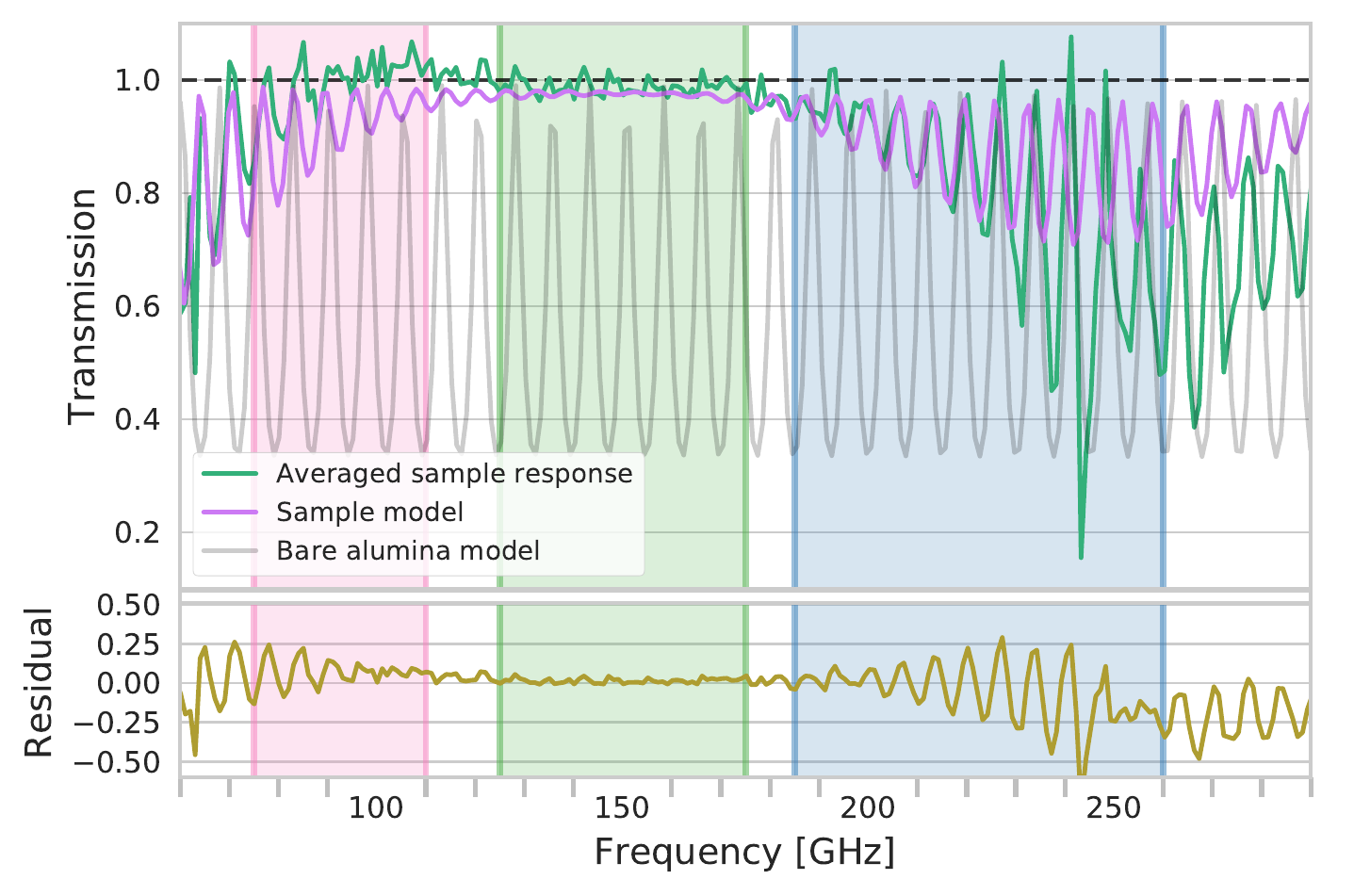}
\end{tabular}
\end{center}
\caption[plots]
{ \label{fig:fts_plot}
AR coating transmission efficiency measured via FTS at $\sim$300 K. The vertical shaded regions demarcate the SPT-3G observing bands centered at 95, 150, and 220 GHz, respectively. Below each set of transmission curves we plot the residuals for that set.
\emph{Left:} The measured response of Sample 1 is shown in blue, the measured response of Sample 2 is shown in red, the predicted response of an uncoated alumina sample is shown in grey. The Sample 1$-$Sample 2 residuals are shown in the lower panel in yellow.
\emph{Right:} The averaged response of Samples 1 and 2 is shown in green, the predicted response of the SPT-3G lenslet AR coating is shown in purple, and the predicted response of an uncoated alumina sample is shown in grey. The average transmission$-$model residuals are again shown in the lower panel in yellow.}
\end{figure}

Two samples were measured, both were alumina pucks (CoorsTek formulation AD-995-I2 -- the same formulation as the lenslets) 2 inch outer diameter by 0.25 inch thick that were coated on both sides by the lenslet AR coating.
The coating response was measured between 50 and 300 GHz at $\sim$300 K.
The results are shown in Fig. \ref{fig:fts_plot}.
In the left panel we show the response of the individual samples; in the right panel we show the averaged response of the two samples and overlay the predicted response.
Table \ref{table:trans_table} lists the mean in-band transmission for Sample 1 and Sample 2, as well as the expected in-band transmission based on the model.
The model takes into account the refractive index, loss tangent, and thickness of each material; it also factors dielectric loss into its output.
We interpret the data below in Sec. \ref{sec:discussion}.

In addition to measuring the transmission response of the coating, we also tested its mechanical robustness to thermal cycling.
Due to the mismatch in the coefficient of thermal expansion of alumina/silicon and the PTFE-based AR coating materials, it is possible for the AR coating to peel away from the optical element when the assembly is cooled to cryogenic temperatures, especially over the course of several thermal cycles.

Early in the research and development process we stress tested the coated lenslet arrays by rapidly submerging the assembly in liquid nitrogen (T$=$77 K), allowing it to thermalize, then removing the assembly from the liquid nitrogen and allowing it to return to room temperature.
The process was then repeated 20 times, or until the coating failed.
We found that if a coating did not peel away from the lenslet array after the first two cycles, then the coating tended to remain anchored for the remaining cycles.
This test is far more extreme than any scenario the lenslet arrays might face in normal operation.
During production of the deployment-grade lenslet arrays, we relaxed the passing conditions of the thermal cycling test; we instead required that the lenslet array survive three slow thermal cycles.

We thermally tested the large-format optics as well.
Due to their size and deployment schedule time constraints, however, we were unable test them as thoroughly as the lenslet arrays.
The large-format optics were similarly submerged in a liquid nitrogen bath and allowed to thermalize.
Each optical element was thermally cycled at least once before deployment and mechanical clamps were added around the perimeter of each lens to prevent coating delamination.

\begin{table}[t]
\caption{Mean transmission in each SPT-3G observing band for the FTS measurements of Sample 1 and Sample 2, as well as the expected transmission based on the modeled sample and modeled bare alumina.}
\label{table:trans_table}
\begin{center}
\begin{tabular}{lccc}
\hline
\multirow{1}{*}{AR Coating} & \multirow{1}{*}{95 GHz} & \multirow{1}{*}{150 GHz} & \multirow{1}{*}{220 GHz} \\
\hline
Sample 1             &  0.990  &  0.994  &  0.843  \\
Sample 2             &  0.997  &  0.982  &  0.770  \\
Model                &  0.925  &  0.975  &  0.869  \\
None (bare alumina)  &  0.584  &  0.584  &  0.574  \\
\hline
\end{tabular}
\end{center}
\end{table}

\begin{figure}[t]
\begin{center}
\begin{tabular}{cc}
\includegraphics[width=9cm]{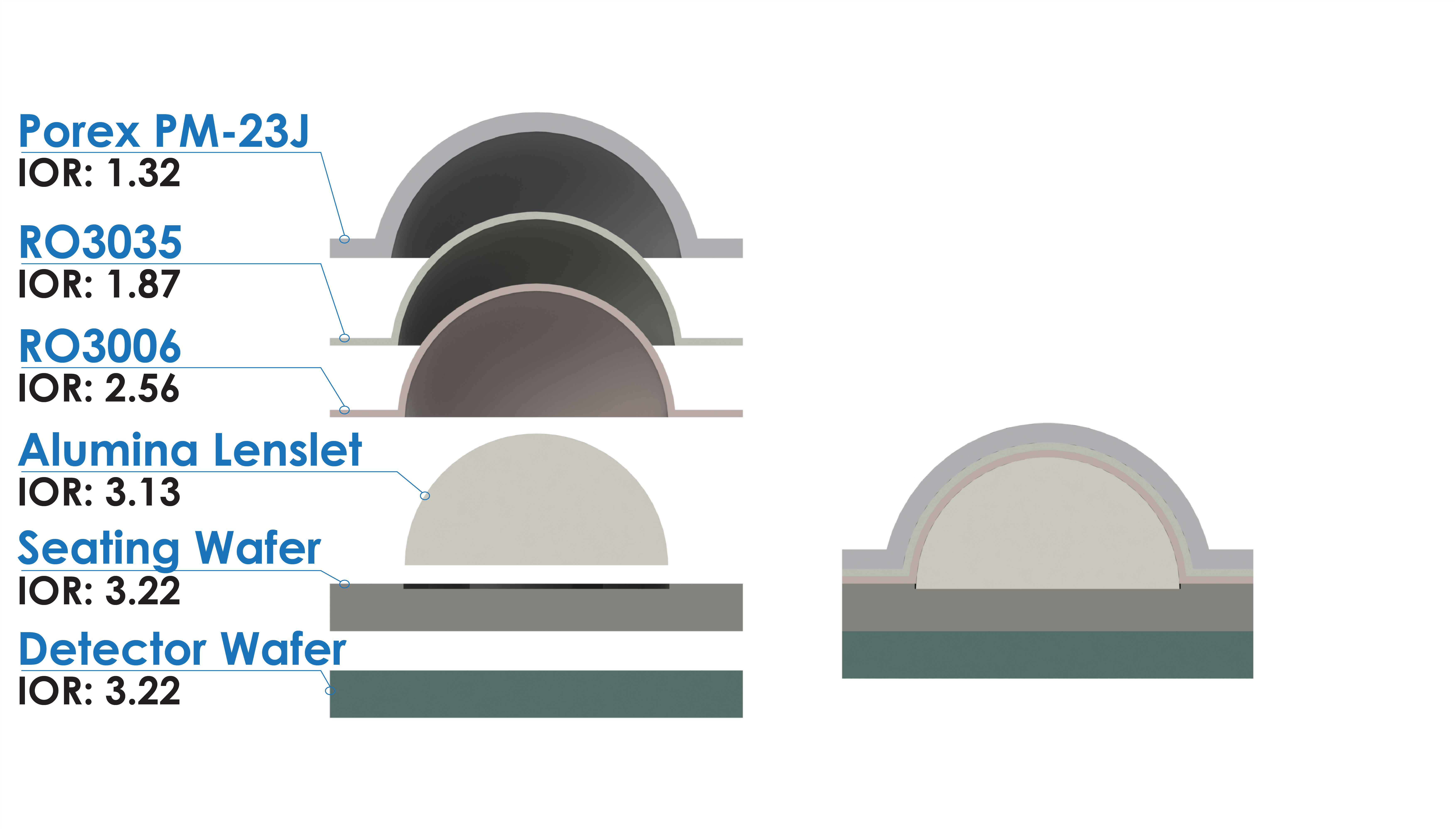}
\includegraphics[height=4.75cm]{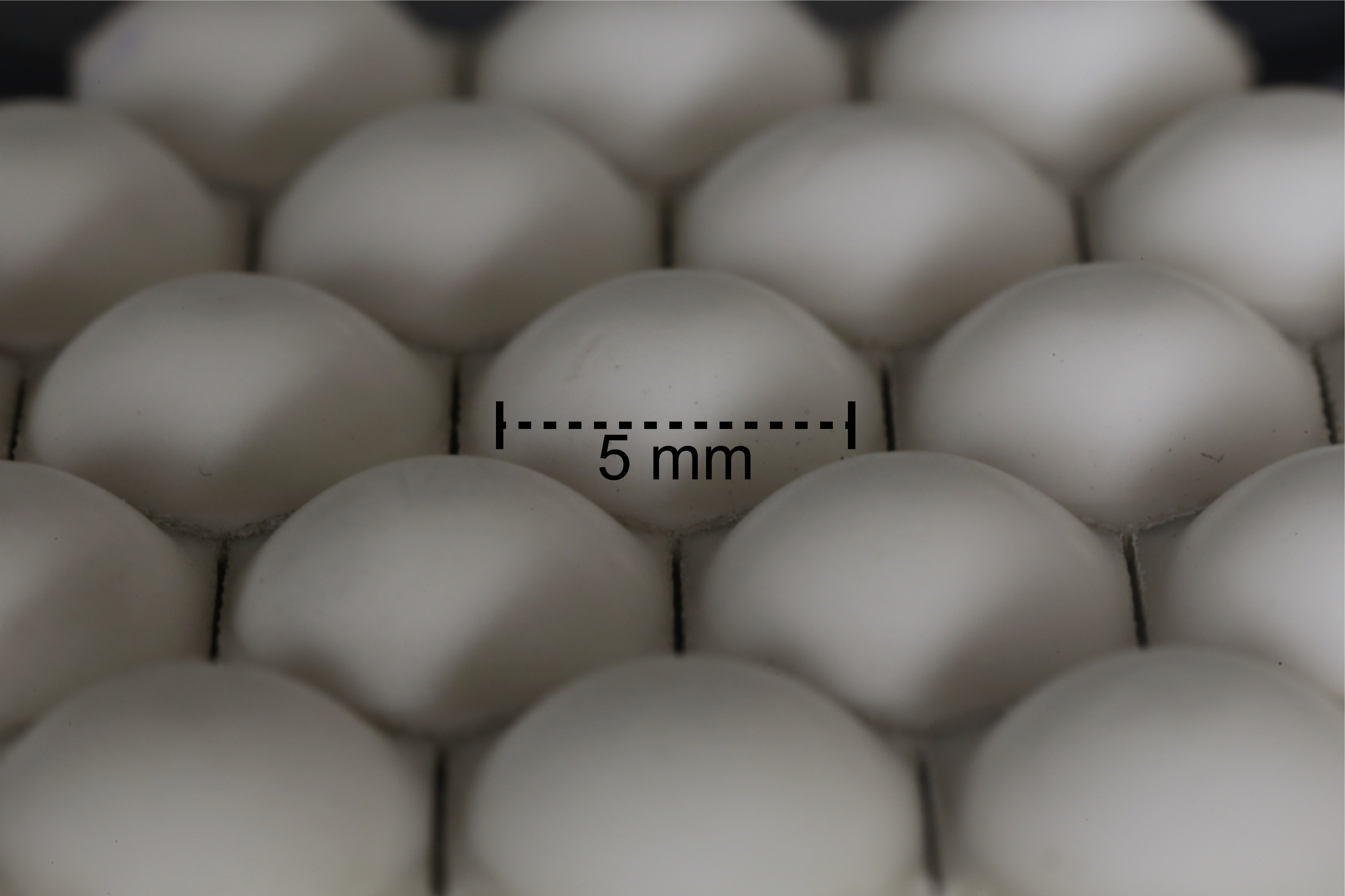} \\
\includegraphics[height=5.5cm]{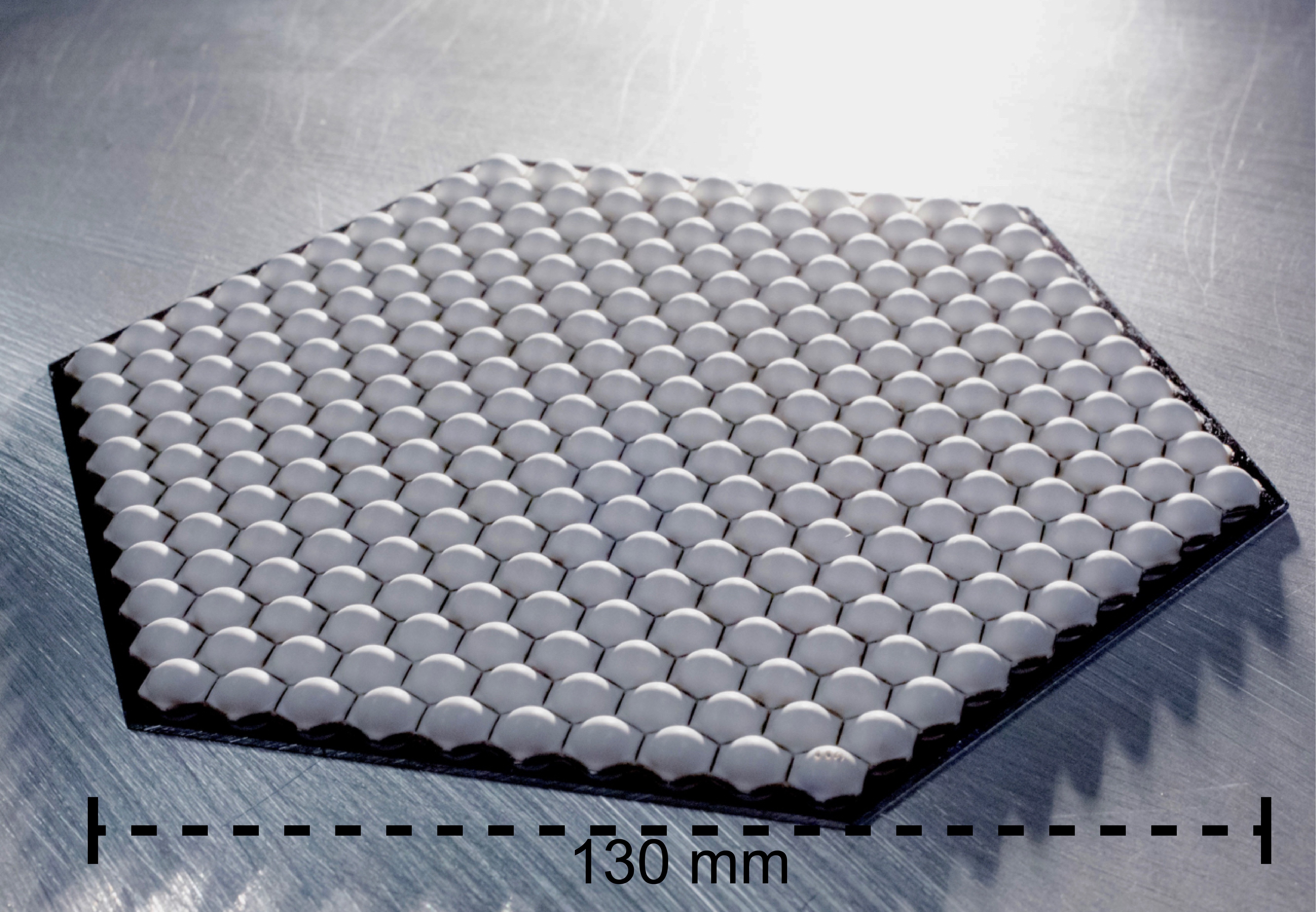}
\includegraphics[height=5.5cm]{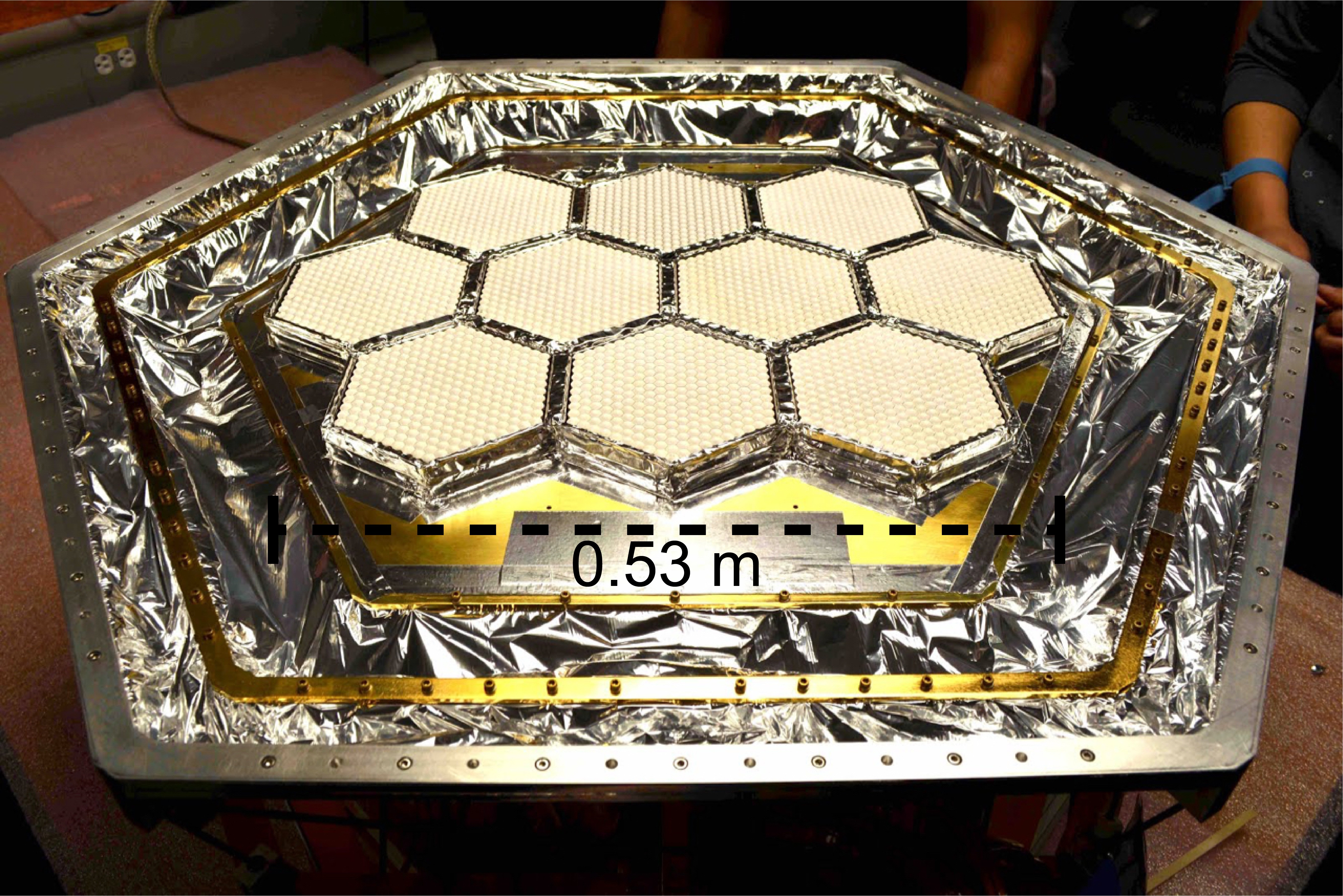}
\end{tabular}
\end{center}
\caption[example]
{ \label{fig:lenslet_array}
(Scales are approximate)
\emph{Top left:} An exploded-view schematic of the AR coating, lenslet, lenslet seating wafer, and detector wafer with the index of refraction (IOR) called out for each material.
\emph{Top right:} A close-up photograph of a lenslet array that has been assembled, AR coated, and laser diced. The high curvature and dense packing of the lenslets is prominently featured.
\emph{Bottom left:} A photograph of an AR coated lenslet array. Each lenslet array consists of 271 hemispherical alumina lenslets mounted to a silicon substrate. The hexagonal silicon substrate is cut from a 6 in diameter silicon wafer.
\emph{Bottom right:} A photograph of the SPT-3G focal plane taken just before its final integration with the rest of the instrument. The ten white hexagons are AR coated lenslet arrays.}
\end{figure}

\section{DISCUSSION}
\label{sec:discussion}
We have developed a three-layer AR coating for the SPT-3G camera's large-format optics and focal plane (Figs. \ref{fig:large_lens} and \ref{fig:lenslet_array}, repsectively).
The lenslet array AR coating is robust to thermal cycling; the large-format coating also survives thermal cycling.
While both coatings improve the overall transmission in the SPT-3G science bands, the coating response deviates from what we expect based on our model, which underestimates transmission in the 95 and 150 GHz bands and overestimates it in the 220 GHz band.
These differences, however, are not entirely surprising.
We use as input to our simulations commercial product specifications for the index of refraction and loss tangents of each material, but these parameters are known to be both frequency- and temperature-dependent and the values provided by manufacturers are derived from measurements made at frequencies below those of our observing bands.
We are in the process of preparing a Univeristy of Chicago-based Michelson FTS for the purpose of measuring each AR coating material individually.

We do not expect, however, that whatever discrepancies we find between the lower-frequency specification values and the higher-frequency measurements will explain all of the model residuals.
Of particular interest to us is the averaged sample response's dip in transmission above $\sim$200 GHz as compared to the model.
It is unlikely that this dip is due entirely to dielectric loss.
For instance, in the right panel of Figure \ref{fig:fts_plot}, we see that beginning at about 220 GHz the peaks in the model curve fall out-of-sync with the averaged response curve.
This shifting is indicative of reflections between the AR coating layers and the alumina.
Additionally, based on the model, we expect a roughly three percent decrease in transmission due to loss in the 220 GHz band.
While this may be consistent with the 220 GHz band mean transmission of Sample 1, that sample's response is artificially inflated due to an imperfect reference measurement (in general, the fractional transmission should not exceed one).
The true value probably lies closer to the 220 GHz band mean transmission of Sample 2, which becomes harder to explain through dielectric loss alone.
A working hypothesis is that we see excess scattering with increasing frequency due to the wavelength of the light approaching the size of the high-index dopants added to the RO3035 and RO3006 layers; we are currently exploring that mechanism.

Even if the model perfectly matched the measured response, there is still room to improve transmission, especially in the 220 GHz band.
Because total transmission, $T$, is exponentially dependent on the number of optical surfaces, $N$, increasing transmission by any fraction is a worthwhile endeavor: $T \propto a^{N}$, where $a$ is the transmission through a single optical surface (assuming all surfaces are identical).
While it is difficult to change the refractive index of the AR coating materials, we can adjust their thicknesses.
By optimizing the thickness of each layer we can increase transmission across all three science bands.
We are currently exploring techniques (e.g., industrial roll forming processes) to accurately and consistently adjust the thickness of the AR coating materials.

We are in the early stages of research and development for a 30 to 40 GHz AR coating.
While the fielded coating is lossy above 200 GHz, there is little loss in the 30 and 40 GHz bands -- bands that are of particular interest to the upcoming CMB Stage-4 (CMB-S4) experiment \cite{cmbs4_science2016, cmbs4_technology2017}.
Preliminary simulations indicate that a PTFE-based coating can achieve a flat response at $>$95\% transmission in the $\sim$25 to $\sim$45 GHz band.
We have constructed a prototype of such a coating using the same materials and techniques described above, and intend to test the coating's response later this year.
Owing to the thickness of the coating, it more readily lends itself to large-format optics rather than high-curvature lenslets.

\section{CONCLUSION}
\label{sec:conclusion}

We have developed a PTFE-based AR coating that is broadband, cryogenically stable, and inexpensive to manufacture -- the cost of the coating materials is $\sim$\$960 per square meter.
The cost of coating a six inch lenslet array is $\sim$\$20 and the cost of coating an entire large-format optic (approximately 720 mm outer diameter) is $\sim$\$1250.
The raw materials are commercially available and the coating can be fabricated in-house, leading to fast design iteration.
The coating is adaptable to both high-curvature lenslets and large diameter wide-field optics.
In addition, the coating has been field-tested -- ten AR coated lenslet arrays were deployed with the SPT-3G experiment in the 2016-2017 austral summer and AR coated wide-field optics were deployed in the 2017-2018 austral summer.

The lenslet coating technique is scalable and several steps can be automated, further reducing manufacturing time.
Scalability and automation are attractive assets as the ground-based CMB research field begins to prepare for the near-future CMB-S4 experiment.
Furthermore, the materials and techniques described above can be applied to lower-frequency scientifically interesting bands.

\acknowledgments
The South Pole Telescope program is supported by the National Science Foundation through grant PLR1248097.
Partial support is also provided by the NSF Physics Frontier Center grant PHY-0114422 to the Kavli Institute of Cosmological Physics at the University of Chicago, the Kavli Foundation, and the Gordon and Betty Moore Foundation through grant GBMF\#947 to the University of Chicago.
This work is also supported by the U.S. Department of Energy.

This material is based upon work supported by the National Science Foundation Graduate Research Fellowship under Grant No. DGE - 1144245.
Work at Argonne National Lab is supported by UChicago Argonne LLC, Operator of Argonne National Laboratory (Argonne).
Argonne, a U.S. Department of Energy Office of Science Laboratory, is operated under contract no. DE-AC02-06CH11357.
We also acknowledge support from the Argonne Center for Nanoscale Materials.
Vieira acknowledges support from the Sloan Foundation.

\bibliography{ar}
\bibliographystyle{spiebib}

\end{document}